# Soliciting opinions and solutions on the " Q Zhang's Problem "

Zhong Wang

(Beijing Doers Education Consulting Co., Ltd)

**Abstract:** "Q Zhang's Problem " is a teaching problem proposed by Qian Zhang, a science teacher at Dongjiao Minxiang Primary School in Dongcheng District, Beijing. In 2022, she proposed that: (1) when explaining the knowledge points of frequency in the "Sound" unit of textbook, experiments on the vibration of objects such as rubber bands and steel rulers were used to assist students in learning, but the effect was not obvious, because it was difficult for the naked eye to distinguish the speed of vibration of objects such as rubber bands, and it was difficult to correspond to the high and low frequencies; (2) Students seem to be confused about the difference between frequency and amplitude. When guiding them to make the rubber band vibrate faster, they tend to tug harder at the rubber band, but this actually changes the amplitude rather than the frequency (changing the frequency should be to control its vibrating chord length, similar to the tuning method of a stringed instrument). Therefore, demonstration experiments using objects such as rubber bands as frequencies do not seem suitable and cannot effectively assist students in establishing the concept of frequency. We hope to solicit opinions and solutions (research ideas) on this problem, with a focus on two points: ① the mathematical/physical explanation of the problem. That is, does simply changing the amplitude really not affect the original vibration frequency of the object (except when the amplitude is 0) ② explanation from a cognitive perspective: Why do people confuse the two concepts? What is the cognitive mechanism behind it.

## 1. Introduction

"Q Zhang's Problem " is a teaching problem proposed by Qian Zhang, a science

teacher at Dongjiao Minxiang Primary School in Dongcheng District, Beijing. In 2022, she proposed that: (1) when teaching the "frequency" concept in the "Sound" unit, experiments on the vibration of objects such as rubber bands and steel rulers were used to assist students in learning, but the effect was not obvious, because it was difficult for the naked eye to distinguish the speed of vibration of objects such as rubber bands, so it was difficult to correspond to the high and low frequencies; (2) Students seem to be confused about the difference between frequency and amplitude. When guiding them to make the rubber band vibrate faster, they tend to tug harder at the rubber band, but this actually changes the amplitude rather than the frequency (changing the frequency should be to control its vibrating chord length, similar to the tuning method of a stringed instrument). Therefore, experiments using objects such as rubber bands as frequencies do not seem suitable and cannot effectively assist students in establishing the concept of frequency.

Because rubber bands and steel rulers are easily available to students, and they have advantages such as high vibration frequencies, ease of hearing vibration sounds, and ease of adjustment, establishing frequency concepts through these experiments is a widely used experimental method, not only in the textbook to which this lesson belongs (Hunan Science and Technology Press), but also in other versions of textbooks. Therefore, the proposal of this problem has also triggered a resonance among teachers using other versions of textbooks. They have all reported that the experiment has a tendency to cause students to confuse amplitude and frequency, indicating that this problem has a certain universality.

## 2. Existing viewpoints

From the literatures, teaching discussions on frequency are more secondary school content, as they require quantitative research using mathematical knowledges. For example, Shunyao Zhang used a vernier caliper and a measuring cylinder to measure the inner volume of a wine bottle, supplemented by audio analysis software to determine the frequency of the sound by blow the wine bottle, and then guided students to study the quantitative relationship between frequency and the volume of air in the bottle (Shunyao Zhang, 2018). Another example is Bangzhen Pan, who worked out the "12-equal temperament" scale frequency relationship through frequency ratio, corrected another paper, and made suggestions for the compilation of physics textbooks (Bangzhen Pan, 1990). In terms of experiments, secondary schools also generally follow similar experiments of Q Zhang's problem. It is worth noting that Yongpin Gao proposed to use sound to tilt the candle flame, and then observe the tilt degree of the flame to determine the strength of frequency, which is relatively special in many experiments (Figure 1) (Yongpin Gao, 2009).

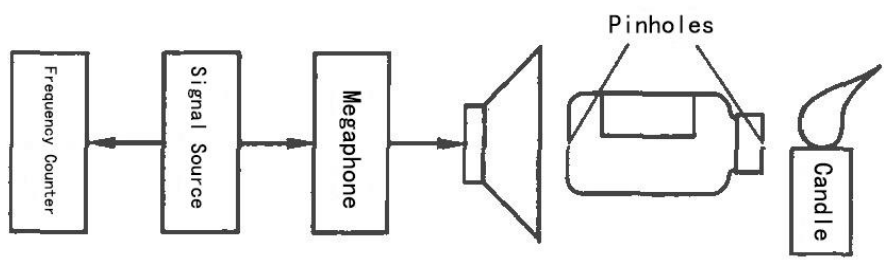

Figure 1

Interestingly, the issue of the relationship between frequency and amplitude in vibration systems has also attracted the attention and discussion of physics teachers. For example, in the article "Is the vibration energy independent of frequency?", Xuefeng Li obtained the total energy formula of the system by analyzing the motion of a spring oscillator and a simple harmonic pendulum:

$$E = \frac{2\pi^2 mA^2}{T^2} = 2\pi^2 mf^2A^2$$

(where m is the mass of the oscillator/pendulum, f is the frequency, T is the period, and A is the amplitude). From this, it is proposed that the total vibration energy of the system is related to both amplitude and frequency (Xuefeng Li, 2002). Keming Tang found through the analysis of a spring oscillator that "the amplitude of forced vibration reach its peak only when the driving frequency is exactly equal to its natural frequency as expected" is not correct when the oscillator vibrate damping, it exhibits an obvious phenomenon of being affected by damping energy (Keming Tang, 2003). However, throughout the literature, there is no research on the issues discussed in this article, especially the confusion of amplitude and frequency.

Therefore, in October 2022, we organized a discussion on this issue (Zhong Wang et al., 2022). Participants' views are basically divided into two schools: improving existing experiments and questioning the experiment.

Teachers who hold the former view mainly hope to use other methods/materials to replace rubber bands and steel rulers as demonstration experiments. As proposed by Zhen Cui, the phenomenon of tuning and amplifying the vibration of the guitar strings and complemented by an explanation of the concept of guitar "fret" to promote the establishment of the concept of frequency (Zhen Cui, 2022). Another example is Yi Zhang 's proposal to use a flashlight and a curtain to amplify the phenomenon of tuning fork vibration, thereby facilitating students' observation. He also proposes a teaching approach that combines the history of science and technology with the history of music, using the research results of ancient scholars on the relationship between music and frequency, such as the "Pythagorean Intonation", as the material. By explaining the exploration process of these scholars on the relationship between music and frequency, students are guided to understand the role of frequency in it, thereby promoting the construction of concepts (Yi Zhang, 2022).

Zhong Wang and others hold another view. It believes that frequency is essentially another representation of periodic phenomena, and therefore teaching the "frequency" concept should be beneficial to students' future construction of the concept of periodicity. But all above experiments are difficult to touch on this aspect. At best, these teaching methods can only leave students with fragmented knowledge that" some methods can be designed to change pitch. " However, due to the cognitive characteristics of children, it is difficult to conduct in-depth guidance and correction on the correctness of the methods designed by students (that is, it is difficult to explain to students why the method of tugging at the rubber band is incorrect). This provides theoretical possibilities and breeding grounds for conceptual confusion. However, the periodic phenomenon is secondary school knowledge, which significantly exceeds children's cognitive level and knowledge accumulation, and is therefore difficult to guide and penetrate (because it is not a purely scientific subject problem, it including mathematical knowledges). Therefore, his suggestion of "cold treatment" essentially implies a question as to whether it is necessary to explain this part of the content at the primary school stage (i.e., the current curriculum standard "Curriculum Content" 3.3, Ministry of Education of the PRC, 2022, P36).

However, we believe that the above discussion and its views do not substantially solve the problem, and there is a certain bias in the direction of the views, which is highlighted in two points: First, the discussion mainly focuses on the methods, namely, is it appropriate to use methods such as rubber bands? Is there a better alternative? But, the discussion of the reason of the problem - why rubber band experiments are prone to conceptual confusion - is insufficient. However, if we do not understand the cause of the problem and only rely on visual inspection of the effectiveness of teaching improvement, it is difficult to say that many of the above alternative experiments are effective and can solve the problem.

Secondly, although the discussion attempted to explore the possible causes of the problem, including citing evidence from cognitive science and cognitive neuroscience, there is a bias in the focus of these evidences: that is, they all point to how children establish abstract concepts. However, this can only serve as the cognitive basis and premise for conceptual confusion, and the formation of erroneous abstract concepts does not necessarily lead to confusion. In other words, the focus of the cognitive perspective on this issue should be: Why do people confuse the two concepts?

## 3. Solicitation requirements

Therefore, we solicit different perspectives and solutions (research assumptions) for this problem. We welcome any views that contribute to solving this problem, but we focus on the following two points:

① the mathematical/physical explanation of the problem. That is, does simply changing the amplitude really not affect the vibration frequency of the object (except when the amplitude is 0)?

② explanation from a cognitive perspective: Why do people confuse the two concepts? What is the cognitive mechanism behind it?

If you are interested in this problem, please organize your views, opinions, and solutions to the problem into a WORD document and send it as an attachment to the email address: **tougao@bdice.ac.cn** (If there are pictures attached, please transcode them into PDF format and send them together with the WORD document).

This collection is a long-term activity. If there are a certain number of ideas and opinions on this content, we will publish a compilation and share it with people.